\documentclass[conference]{IEEEtran}
\IEEEoverridecommandlockouts
\usepackage{cite}
\usepackage[ruled,vlined]{algorithm2e}
\usepackage{amsmath,amssymb,amsfonts}
\usepackage{algorithmic}
\usepackage{graphicx}
\usepackage{textcomp}
\usepackage{xcolor}
\def\BibTeX{{\rm B\kern-.05em{\sc i\kern-.025em b}\kern-.08em
    T\kern-.1667em\lower.7ex\hbox{E}\kern-.125emX}}

    \renewcommand{\baselinestretch}{0.95}

\begin{document}

\title{PARAFAC-Based Channel Estimation for Intelligent Reflective Surface Assisted MIMO System\\
}

\author{\IEEEauthorblockN{Gilderlan T. de Ara\'{u}jo}
\IEEEauthorblockA{\textit{Federal Institute of Cear\'{a}}\\
Canind\'{e}, Brazil \\
gilderlan.tavares@ifce.edu.br}
\and
\IEEEauthorblockN{Andr\'{e} L. F. de Almeida}
\IEEEauthorblockA{\textit{Federal University of Cear\'{a}}}
Fortaleza, Brazil \\
andre@gtel.ufc.br}

\maketitle

\begin{abstract}
Intelligent reflective surface (IRS) is an emergent technology for future wireless communications. It consists of a large 2D array of passive scattering elements that control the electromagnetic properties of radio-frequency waves so that the reflected signals add coherently at the intended receiver or destructively to reduce co-channel interference. The promised gains of IRS-assisted communications depend on the accuracy of the channel state information. In this paper, we propose two novel channel estimation methods for an IRS-assisted multiple-input multiple-output (MIMO) communication system. Assuming a structured time-domain pattern of pilots and IRS phase shifts, we show that the received signal follows a parallel factor (PARAFAC) tensor model that can be exploited to estimate the involved communication channels in closed-form or iteratively. Numerical results corroborate the effectiveness of the proposed channel estimation methods and highlight the involved tradeoffs.
\end{abstract}


\section{Introduction}
Recently, we have experienced an increase in the expectation of the effective use of the 5G technology. However, researchers are already thinking about the potential technologies for beyond 5G wireless communications. Intelligent reflective surface (IRS) (also referred to as reconfigurable intelligent surface or software-controlled metasurface) \cite{Basar_2019, Survey_NOVO, Liaskos2018ANW, Jung2019, Huang2019,basar2019reconfigurable} consists of a large 2D array composed of nearly passive, low- cost, reflecting elements whose parameters are adjusted so that the reflected signals add coherently at the intended receiver or destructively to reduce co-channel interference. Each element can act independently and can be reconfigured in a software-defined manner by means of an external controller. The IRS does not require dedicated radio-frequency chains and is usually wirelessly powered by an external RF-based source, as opposed to amplify-and-forward or decode-and-forward relays, which require dedicated power sources \cite{Huang2019}.

Several recent works have discussed the potentials and challenges of IRS-assisted wireless communications (see, e.g., \cite{Basar_2019},\cite{Survey_NOVO} and references therein). Among the several open issues, we highlight the acquisition of channel state information. Two main obstacles for channel estimation in IRS-assisted communications are the passive nature and the large number of IRS elements. Recently, a few works have addressed the channel estimation problem and provided different solutions. In \cite{Tobias2019}, a minimum variance unbiased estimator is proposed, and an optimal design of the IRS phase shift matrix is found. The authors of \cite{ZHEN2019} propose a two-stage algorithm by exploiting sparse representations of low-rank multipath channels. In \cite{NING2019}, links between massive MIMO and IRS are discussed in the context of Terahertz communications, and a cooperative channel estimation via beam training is presented. In \cite{Yaoshen2019}, IRS is proposed as a solution to mitigate the blockage problem in mmWave communications and a channel estimation approach is presented. In \cite{Emil:NOV2019}, channel estimation is carried out by resorting to an on-off strategy that sequentially activates the IRS elements one-by-one. 

In this paper, we establish a connection between IRS-assisted MIMO communication and tensor decomposition. By assuming a structured time-domain pattern of pilots and IRS phase shifts, we show that the received signal follows a parallel factor (PARAFAC) tensor model. Exploiting its algebraic structure, we propose two simple and effective algorithms to estimate the transmitter-IRS and IRS-receiver MIMO channels, respectively. The first algorithm is a closed-form solution based on rank-1 matrix approximations, while the second consists of an iterative bilinear alternating least squares algorithm. While the first algorithm is an algebraic and less complex, the second one can operate under less restrictive conditions on the training parameters. Illustrative numerical results are provided to evaluate the performance of the proposed channel estimation methods.


\

\vspace{-2ex}
\noindent \textit{Notation:} Matrices are represented with boldface capital letters ($\mathbf{A}, \mathbf{B}, \dots)$, and vectors are denoted by boldface lowercase letters ($\mathbf{a}, \mathbf{b}, \dots)$. Tensors are symbolized by calligraphic letters $(\mathcal{A}, \mathcal{B}, \dots)$. Transpose and pseudo-inverse of a matrix $\mathbf{A}$ are denoted as $\mathbf{A}^T$ and $\mathbf{A}^\dagger$, respectively. The operator $\textrm{diag}(\mathbf{a})$ forms a diagonal matrix out of its vector argument, while $\diamond$ and $\otimes$ denote the Khatri Rao and Kronecker products, respectively. $\mathbf{I}_N$ denotes the $N \times N$ identity matrix. The operator $\textrm{vec}(\cdot)$ vectorizes an $I \times J$ matrix argument, while $\textrm{unvec}_{I \times J}(\cdot)$ does the opposite operation. 

\section{System Model}
\begin{figure}[!t]
	\centering\includegraphics[scale=0.5]{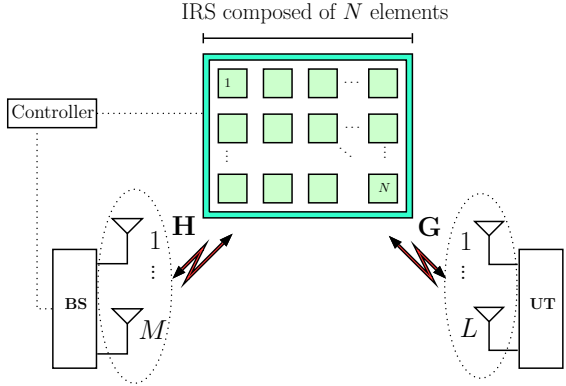}
	\vspace{-2.5ex}
	\caption{IRS-assisted  MIMO system}
	\label{fig:LIS}
\end{figure}

We consider a MIMO communication system assisted by an IRS. Both the transmitter and the receiver are equipped with multiple antennas. Although the terminology adopted in this paper assumes a downlink communication, where the transmitter is the base station (BS) and the receiver is the  user terminal (UT), our signal model also applies to the uplink case by just inverting the roles of the transmitter and the receiver. The base station and user terminal are equipped with arrays of $M$ and $L$ antennas, respectively. The IRS is composed of $N$ elements, or unit cells, capable of individually adjusting their reflection coefficients (i.e., phase shifts). The line-of-sight (LOS) path between the BS and UT is assumed to be unavailable. 
The system model is illustrated in Figure \ref{fig:LIS}. Assuming a block-fading channel, the received signal model is usually given as follows \cite{ZHEN2019}
\begin{equation}
\mathbf{y}[t] = \mathbf{G}(\mathbf{s}[t] \odot \mathbf{H}\mathbf{x}[t]) + \mathbf{n}[t], \quad t = 1, \dots, T,
\label{Eq:EquationFromRef}
\end{equation}
or, alternatively,
\begin{equation}
\mathbf{y}[t] = \mathbf{G}\textrm{diag}(\mathbf{s}[t])\mathbf{H}\mathbf{x}[t] + \mathbf{n}[t],
\label{Eq:EquationFromRef2}
\end{equation}
where $\mathbf{x}[t] \in \mathbb{C}^{M \times 1}$ is the vector containing the transmitted pilot signals at time $t$, $\mathbf{s}[t] = \left[s_{1,t}e^{j\phi_1},\dots,s_{N,t}e^{j\phi_N}\right]^T \in \mathbb{C}^{N \times 1}$  is the vector that models the phase shifts and activation pattern of the IRS, where $\phi_n \in (0 , 2\pi]$, and $s_{n,t} \in \{0 , 1\}$ controls the on-off state of the corresponding element at time $t$. The matrices $\mathbf{H} \in \mathbb{C}^{N \times M}$ and $\mathbf{G} \in \mathbb{C}^{L \times N}$ denote the BS-IRS and IRS-UT MIMO channels, respectively, while
$\mathbf{n}[t] \in \mathbb{C}^{L \times 1}$ is the additive white Gaussian noise (AWGN) vector. We assume that the entries of the BS-IRS and IRS-UT channel matrices $\mathbf{H}$ and $\mathbf{G}$ are independent and identically distributed zero-mean circularly-symmetric complex Gaussian random variables. 

\begin{figure}[!t]
	\centering\includegraphics[scale=0.9]{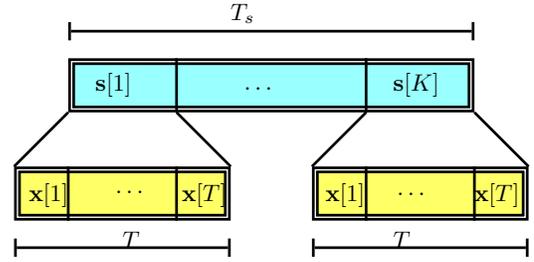}
	\vspace{-1ex}
	\caption{Structured pilot pattern in the time domain}
	\label{fig:bockdiagram}
\end{figure}

The channel coherence time $T_s$ is divided into $K$ blocks, where each block has $T$ time slots so that $T_s = KT$. Let us define $\mathbf{y}[k,t]\doteq y[(k-1)T + t]$ as the received signal at the $t$-th time slot of the $k$-th block, $t=1,\dots, T$, $k=1, \dots, K$. Likewise, denote $\mathbf{x}[k,t]$ and $\mathbf{s}[k,t]$ as the pilot signal and phase shift vectors associated with the $t$-th time slot of the $k$-th block. We propose the following structured time-domain protocol: i) the IRS phase shift vector is constant during the $T$ time slots of the $k$-th block and varies from block to block; ii) the pilot signals $\{\mathbf{x}[1], \dots, \mathbf{x}[T]\}$ are repeated over the $K$ blocks. Mathematically, this means that
\begin{eqnarray}
&\mathbf{s}[k,t] = \mathbf{s}[k], \,\, \text{for}\,\, t = 1, \dots, T,\\
&\mathbf{x}[k,t] = \mathbf{x}[t], \,\, \text{for}\,\, k = 1, \dots, K.
\end{eqnarray}
An illustration of this time-domain protocol is shown in Figure \ref{fig:bockdiagram}. Under these assumptions, the received signal model (\ref{Eq:EquationFromRef2}) can be written as
\label{sysmodel}

\begin{equation}
\mathbf{y}[k,t] = \mathbf{G}\textrm{diag}(\mathbf{s}[k])\mathbf{H}\mathbf{x}[t] + \mathbf{n}[k,t],
\label{Eq:ReceivedVector}
\end{equation}
Collecting the received signals during $T$ time slots for the $k$-th block in $\mathbf{Y}[k] = \left[\mathbf{y}[k,1] \dots \mathbf{y}[k,T]\right] \in \mathbf{C}^{L \times T}$ leads to
\begin{equation}
\mathbf{Y}[k] = \mathbf{G}\mathbf{D}_k(\mathbf{S})\mathbf{H}\mathbf{X}^\textrm{T} + \mathbf{N}[k],
\label{Eq:PARAFAC_MODEL}
\end{equation}
where $\mathbf{X}\doteq [\mathbf{x}[1], \dots, \mathbf{x}[T]]^{\textrm{T}} \, \in \mathbb{C}^{T \times M}$, $\mathbf{N}\doteq [\mathbf{n}[1], \dots, \mathbf{n}[T]]^{\textrm{T}} \, \in \mathbb{C}^{L \times T}$, $\mathbf{S}\doteq [\mathbf{s}[1], \dots, \mathbf{s}[K]]^{\textrm{T}} \, \in \mathbb{C}^{K \times N}$,
and $\mathbf{D}_k(\mathbf{S}) \doteq \textrm{diag}(\mathbf{s}[k])$ denotes a diagonal matrix holding the $k$-th row of the IRS phase shift matrix $\mathbf{S}$ on its main diagonal.

In order to simplify the exposition of the signal model, we remove the noise term from the following developments. The noise term will be taken into account later. 
We can rewrite the signal part of equation (\ref{Eq:PARAFAC_MODEL}) as
\begin{equation}
\overline{\mathbf{Y}}[k] = \mathbf{G}\mathbf{D}_k(\mathbf{S}) \mathbf{Z}^\textrm{T}, \quad \mathbf{Z} \doteq \mathbf{X}\mathbf{H}^\textrm{T} \in \mathbb{C}^{T \times N}.
\label{eq:PARAFAC MODEL}	
\end{equation}
The matrix $\overline{\mathbf{Y}}[k]$ can be viewed as the $k$-th frontal matrix slice of a three-way tensor $\overline{\mathcal{Y}} \in \mathbb{C}^{L \times T \times K}$ that follows a PARAFAC decomposition, also known as canonical polyadic decomposition (CPD) \cite{Harshman70,Kolda2009,CLdA09,Almeida2016,Sidiropoulos2017}). Each $(\ell,t,k)$-th entry of the noiseless received signal tensor $\overline{\mathcal{Y}}$ can be written as:
\begin{equation}\label{scalar_Y}
[\overline{\mathcal{Y}}]_{\ell,t,k}= \sum\limits_{n=1}^N g_{\ell,n}z_{t,n}s_{k,n},
\end{equation}
where $g_{\ell,n}\doteq [\mathbf{G}]_{\ell,n}$, $z_{t,n}\doteq [\mathbf{Z}]_{t,n}$, and $s_{k,n}\doteq [\mathbf{S}]_{k,n}$. A shorthand notation for the PARAFAC decomposition (\ref{scalar_Y}) is denoted as $\overline{\mathcal{Y}}=[[\mathbf{G}, \mathbf{Z}, \mathbf{S}]]$.
Exploiting the trilinearity of the PARAFAC decomposition,
we can ``unfold'' received signal tensor $\overline{\mathcal{Y}}$ in the following three matrix forms \cite{Harshman70,Kolda2009}:
\begin{eqnarray}
& \overline{\mathbf{Y}}_1= \mathbf{G}(\mathbf{S} \diamond \mathbf{Z})^{\textrm{T}} \,\, \in \mathbb{C}^{L \times TK},\label{unfolding1}\\
&\overline{\mathbf{Y}}_2= \mathbf{Z}(\mathbf{S} \diamond \mathbf{G})^{\textrm{T}} \,\, \in \mathbb{C}^{T \times LK},\label{unfolding2}\\
&\overline{\mathbf{Y}}_3= \mathbf{S}(\mathbf{Z} \diamond \mathbf{G})^{\textrm{T}}\,\, \in \mathbb{C}^{K \times LT},\label{unfolding3}
\end{eqnarray}
where $\overline{\mathbf{Y}}_1\doteq [\overline{\mathbf{Y}}[1], \dots, \overline{\mathbf{Y}}[K]]$, $\overline{\mathbf{Y}}_2\doteq [\overline{\mathbf{Y}}^{\textrm{T}}[1], \dots, \overline{\mathbf{Y}}^{\textrm{T}}[K]]$, and $\overline{\mathbf{Y}}_2\doteq [\textrm{vec}(\overline{\mathbf{Y}}[1]), \dots, \textrm{vec}(\overline{\mathbf{Y}}[K])]^{\textrm{T}}$. In the following, we formulate two channel estimation methods that exploits the algebraic structure of the PARAFAC model (\ref{scalar_Y}). 

\section{PARAFAC-Based channel estimation}\label{Chanel estmation}
Our goal is to estimate the channel matrices $\mathbf{H}$ (BS-IRS) and $\mathbf{G}$ (IRS-UT) from the received signal tensor given in (\ref{scalar_Y}). Let us define $\mathcal{Y} \doteq \overline{\mathcal{Y}} + \mathcal{N}$ as the noise-corrupted received signal tensor, where $\mathcal{N} \in \mathbb{C}^{L \times T \times K}$ is the additive noise tensor. Likewise, $\mathbf{Y}_i\doteq \overline{\mathbf{Y}}_i + \mathbf{N}_i$, $i=1,2,3$, are the noisy versions of the 1-mode, 2-mode and 3-mode matrix unfoldings (\ref{unfolding1})-(\ref{unfolding3}) of the received signal tensor, and $\mathbf{N}_{i=1,2,3}$ the corresponding matrix unfoldings of the noise.

Unless otherwise stated, both the pilot signal matrix $\mathbf{X}$ and the IRS phase shifts matrix $\mathbf{S}$ are designed as semi-unitary matrices satisfying $\mathbf{X}^{\textrm{H}}\mathbf{X}=\mathbf{I}_{M}$ and $\mathbf{S}^{\textrm{H}}\mathbf{S}=\mathbf{I}_{N}$, respectively. A good choice is to design both $\mathbf{X}$ and $\mathbf{S}$ as truncated discrete Fourier transform (DFT) matrices. The optimal design of the IRS matrix $\mathbf{S}$ is discussed in \cite{jensen2019optimal} for the multiple-input single-output (MISO) case (i.e, for single-antenna users).

\subsection{Closed-form solution}
First, note that we can rewrite the noise-corrupted matrix unfolding (\ref{unfolding3}) as:
\begin{equation}
\begin{aligned}
\mathbf{Y}_3  & = \mathbf{S}(\mathbf{Z} \diamond \mathbf{G})^{\textrm{T}} +  \mathbf{N}_3\\
 & = \mathbf{S}\left(\mathbf{H}^{\textrm{T}} \diamond \mathbf{G}\right)^{\textrm{T}}(\mathbf{X}\otimes \mathbf{I}_L)^{\textrm{T}} + \mathbf{N}_3,
\end{aligned}
\label{unfolding3_2}
\end{equation}
where we have applied the property $(\mathbf{A} \otimes \mathbf{B})(\mathbf{C} \diamond \mathbf{D})=(\mathbf{A}\mathbf{C}) \diamond (\mathbf{B}\mathbf{D})$ to the term $(\mathbf{Z} \diamond \mathbf{G})=(\mathbf{X}\mathbf{H}^{\textrm{T}} \diamond \mathbf{G})$.
A bilinear time-domain filtering is applied at the receiver by exploiting the knowledge of the IRS matrix and the pilot signal matrix. This is achieved by left- and right-filtering of $\mathbf{Y}_3$ as
\begin{equation}
\mathbf{W}^{\textrm{T}}  \doteq \mathbf{S}^{\textrm{H}}\mathbf{Y}_3(\mathbf{X}^{\ast}\otimes \mathbf{I}_L) = \left(\mathbf{H}^{\textrm{T}} \diamond \mathbf{G}\right)^{\textrm{T}} + \tilde{\mathbf{N}}_3,
\label{Khatri_ZG}
\end{equation}
where $\tilde{\mathbf{N}}_3=\mathbf{S}^{\textrm{H}}\mathbf{N}_3(\mathbf{X}^{\ast}\otimes \mathbf{I}_L)$ is the filtered noise term. 
Note that $\mathbf{W} \in \mathbb{C}^{ML \times N}$ is a noisy version of the (Khatri-Rao structured) \emph{virtual MIMO channel} that models the IRS-assisted MIMO transmission.

From equation (\ref{Khatri_ZG}), we propose to estimate $\mathbf{H}$ and $\mathbf{G}$ by solving the following least squares problem
\begin{equation}
\underset{\mathbf{H}, \mathbf{G}}{\min} \left\| \mathbf{W} - \mathbf{H}^T \diamond \mathbf{G}\right\|^2_F\label{lskrf_problem}
\end{equation}
An efficient solution to this problem is given by the least squares Khatri-Rao factorization (LSKRF) algorithm \cite{Kibangou2009},\cite{Roemer2010}. Note that the problem (\ref{lskrf_problem}) can be interpreted as finding estimates of $\mathbf{H}$ and $\mathbf{G}$ that minimize a set of rank-$1$ matrix approximations, i.e.,
\begin{equation}
(\hat{\mathbf{H}},\hat{\mathbf{G}})= \underset{\{\mathbf{h}_n\},\{\mathbf{g}_n\}}{\arg\min}  \sum\limits_{n=1}^N  \left\| \tilde{\mathbf{W}}_n - \mathbf{g}_n\mathbf{h}^{\textrm{T}}_n \right\|^2_F\label{lskrf_problem2},
\end{equation}
where $\tilde{\mathbf{W}}_n\doteq \textrm{unvec}_{L \times M}(\mathbf{w}_n) \in \mathbf{C}^{L \times M}$, while $\mathbf{g}_n \in \mathbf{C}^{L \times 1}$ and $\mathbf{h}^{\textrm{T}}_n \in \mathbf{C}^{1 \times M}$ are the $n$-th column and $n$-th row of $\mathbf{G}$ and $\mathbf{H}$, respectively. The estimates of $\mathbf{g}_n$ and $\mathbf{h}_n$ in (\ref{lskrf_problem2}) can be obtained from the dominant left and right singular vectors of $\tilde{\mathbf{W}}_n$, respectively, $n=1, \dots, N$. Hence, our channel estimation problem translates into solving $N$ rank-1 matrix approximation subproblems, for which several efficient solutions exist in the literature \cite{golub13}. A summary of the algorithm, referred to as LSKRF, is given in Algorithm $1$, where t-SVD denotes a truncated SVD that returns the dominant singular vectors and singular value.

\begin{algorithm}
	\DontPrintSemicolon
	\textbf{Procedure}\\
	\Begin{
			   \textit{Bilinear filtering of $\mathbf{Y}_3$}:\;
			   $\mathbf{W}^{\textrm{T}} \longleftarrow \mathbf{S}^{\textrm{H}}\mathbf{Y}_3(\mathbf{X}^{\ast}\otimes \mathbf{I}_L)$\;
		\For{$n = 1, \dots ,N$}{
			$\tilde{\mathbf{W}}_n \longleftarrow \textrm{unvec}_{L \times M}(\mathbf{w}_n)$\;
			$(\mathbf{u}_1,\mathbf{\sigma}_1,\mathbf{v}_1)\longleftarrow\textrm{t-SVD}(\tilde{\mathbf{W}}_n)$\;
			$\hat{\mathbf{h}}_n \longleftarrow \sqrt{\sigma_1}\mathbf{v}_1^\ast$\;
			$\hat{\mathbf{g}}_n \longleftarrow \sqrt{\sigma_1}\mathbf{u}_1$\;	
			\textbf{end}}
		\textit{Reconstruct} $\hat{\mathbf{H}}$ \textit{and} $\hat{\mathbf{G}}$:\;
		$\hat{\mathbf{H}} \longleftarrow \left[\hat{\mathbf{h}}_1, \dots , \hat{\mathbf{h}}_N\right]^T$\;
		$\hat{\mathbf{G}} \longleftarrow \left[\hat{\mathbf{g}}_1, \dots , \hat{\mathbf{g}}_N\right]$\;
	}
	\caption{Least squares Khatri-Rao factorization (LSKRF)
	}
	\label{PseudocodeLSKRF}
\end{algorithm}

\subsection{Iterative solution}
From the noisy versions of the matrix unfoldings (\ref{unfolding1}) and (\ref{unfolding2}), we can derive an iterative solution based on a bilinear alternating least squares (BALS) algorithm. This algorithm consists of estimating the matrices $\mathbf{G}$ and $\mathbf{H}$ in an alternating way by iteratively optimizing the following two cost functions:
\begin{eqnarray}
&\hat{\mathbf{G}} = \underset{\mathbf{G}}{\arg\min} \,\, \left\|\mathbf{Y}_{1} - \mathbf{G}(\mathbf{S} \diamond \mathbf{X}\mathbf{H}^{\textrm{T}})^{\textrm{T}}\right\|_F^2\label{func costG},\\
&\hat{\mathbf{H}} = \underset{\mathbf{H}}{\arg\min}\,\, \left\|\mathbf{Y}_{2} - \mathbf{X}\mathbf{H}^{\textrm{T}}(\mathbf{S} \diamond \mathbf{G})^{\textrm{T}}\right\|_F^2,
\label{Func costZ}
\end{eqnarray}
the solutions of which are respectively given by
\begin{eqnarray}
&\hat{\mathbf{G}} = \mathbf{Y}_1\left[\left(\mathbf{S} \diamond \mathbf{X}\mathbf{H}^{\textrm{T}}\right)^{\textrm{T}} \right]^\dagger,\label{EstimaG}\\
&\hat{\mathbf{H}}^{\textrm{T}} = \mathbf{X}^\dagger\mathbf{Y}_2\left[\left(\mathbf{S} \diamond \mathbf{G} \right)^{\textrm{T}}\right]^\dagger. \label{EstimaH}
\end{eqnarray}
Under the column-orthogonality assumption for $\mathbf{X}$ and $\mathbf{S}$, the left pseudo-inverses in (\ref{EstimaG}) and (\ref{EstimaH}) can be replaced by lower complexity matrix products (details are omitted here due to limited space). The BALS is summarized in Algorithm \ref{PseudocodeBALS}.
\begin{algorithm}
	\DontPrintSemicolon
	\textbf{Procedure}\\
	\begin{enumerate}
		\item[1:] \textit{Set} $i = 0$ \textit{initialize randomly} $\hat{\mathbf{H}}_{i=0}$;
		\item[2:] $i = i + 1$;
		\item [3:] \textit{Find a least squares estimate of} $\mathbf{G}$:\\
		$$\hat{\mathbf{G}}_{(i)} = \mathbf{Y}_1\left[\left(\mathbf{S} \diamond \mathbf{X}\hat{\mathbf{H}}^{\textrm{T}}_{(i-1)}\right)^{\textrm{T}} \right]^\dagger;$$
		\item[4:]  \textit{Find a least squares estimate of} $\mathbf{H}$:\\
		$$
		\hat{\mathbf{H}}^{\textrm{T}}_{(i)} = \mathbf{X}^\dagger\mathbf{Y}_2\left[\left(\mathbf{S} \diamond \mathbf{G}_{(i)}\right)^{\textrm{T}}\right]^\dagger
		$$
		\item[5:]\textit{Repeat steps} $2$ to $4$ \textit{until convergence.}
	\end{enumerate}
	\caption{Bilinear alternating least squares (BALS)}
	\label{PseudocodeBALS}
\end{algorithm}

The convergence is achieved when $\|e_{(i)} - e_{(i-1)}\|\leq 10^{-6}$, where $e_{(i)} = \|\mathcal{Y} - \hat{\mathcal{Y}}_{(i)}\|_{F}^{2}$ denotes the  reconstruction error computed at the $i$-th iteration, where $\hat{\mathcal{Y}}_{(i)}=[\hat{\mathbf{G}}_{(i)}, \mathbf{X}\hat{\mathbf{H}}^{\textrm{T}}_{(i)}, \hat{\mathbf{S}}]$ is the reconstructed PARAFAC model (c.f. (\ref{Eq:PARAFAC_MODEL}), (\ref{scalar_Y})) obtained from the estimated channel matrices $\hat{\mathbf{G}}_{(i)}$ and $\hat{\mathbf{H}}_{(i)}$ at the end of the $i$-th iteration. 
Despite the iterative nature of the BALS algorithm, only a few iterations are necessary for convergence (usually less than 10 iterations) due to the knowledge of the matrix factor $\mathbf{S}$ that remains fixed during the iterations.

\subsection{Design requirements}
The LSKRF method (Algorithm 1) has a bilinear filtering step as shown in (\ref{Khatri_ZG}) requiring that $\mathbf{S}$ and $\mathbf{X}$ be semi-unitary (or column-orthogonal), which implies $K \geq N$ and $T \geq M$. The BALS method (Algorithm 2) requires that $\left(\mathbf{S} \diamond \mathbf{X}\mathbf{H}^{\textrm{T}}\right) \in \mathbb{C}^{KT \times N}$ and $\left(\mathbf{S} \diamond \mathbf{G}\right) \in \mathbb{C}^{KL \times N}$ have full column-rank, so that the LS problems in (\ref{func costG}) and (\ref{Func costZ}) (resp. steps 3 and 4 of Algorithm 2) admit unique solutions. This implies that $\textrm{min}(KT,KL)\geq N$, or, equivalently, $K\textrm{min}(T,L)\geq N$. Hence, the iterative BALS method has a less restrictive requirement on the number $K$ of blocks necessary for channel training compared to the LSKRF method. On the other hand, the later usually has a lower computational complexity as shown in our numerical results.

\subsection{Ambiguities}
Provided that the above conditions are satisfied, the channel estimates $\hat{\mathbf{G}}$ and $\hat{\mathbf{H}}$, for both LS-KRF and BALS algorithms, are unique up to scalar ambiguities. More specifically, the rows of $\hat{\mathbf{H}}$ and the columns of $\hat{\mathbf{G}}$ are affected by scaling factors that compensate each other, i.e., $\hat{\mathbf{H}}= \boldsymbol{\Delta}_H\mathbf{H}$ and $\hat{\mathbf{G}}  = \mathbf{G}\boldsymbol{\Delta}_G$,
where $\boldsymbol{\Delta}_H\boldsymbol{\Delta}_G=\mathbf{I}_N$. However, these ambiguities disappear when building an estimate of the cascaded (end-to-end) channel $\hat{\mathbf{H}}_{\textrm{c}}\doteq \hat{\mathbf{G}}\hat{\mathbf{H}}^{\textrm{T}}$ of the IRS-assisted MIMO system. Note also that permutation ambiguity does not exist due to the knowledge of the IRS matrix $\mathbf{S}$ at the receiver.


\section{Numerical Results}
In this section, some numerical results are presented. The channel estimation accuracy  is evaluated in terms of the normalized mean square error (NMSE) given by $\textrm{NMSE}(\hat{\mathbf{H}}) = \frac{1}{R}\sum_{r=1}^{R} (\|\mathbf{H}^{(r)} - \hat{\mathbf{H}}^{(r)}\|_F^2/ \|\mathbf{H}^{(r)}\|_F^2)$, where $\hat{\mathbf{H}}^{(r)}$ is the BS-IRS channel estimated at the $r$-th run, and $R$ denotes the number of Monte Carlo runs. The same definition applies for the estimated IRS-UT channel. The SNR (in dB) is defined as $\textrm{SNR} = 10\textrm{log}_{10}(\|[\overline{\mathcal{Y}}]\|_F^2/ \|[\mathcal{N}]\|_F^2)$.
\begin{figure}[!t]
\begin{center}
	\includegraphics[scale=0.5]{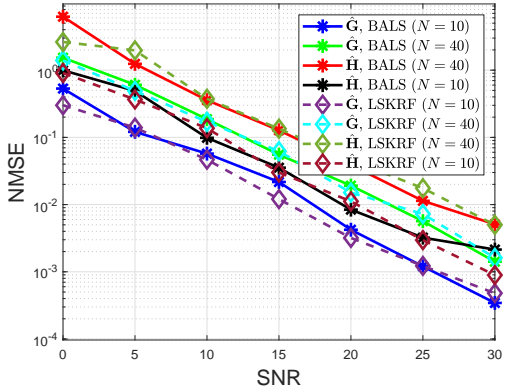}
\end{center}
	\vspace{-3ex}
\caption{NMSE of the estimated channels $\hat{\mathbf{H}}$ and $\hat{\mathbf{G}}$.}
	\label{GHCOMP}
\end{figure}
\begin{figure}[!t]
\begin{center}
	\includegraphics[scale=0.5]{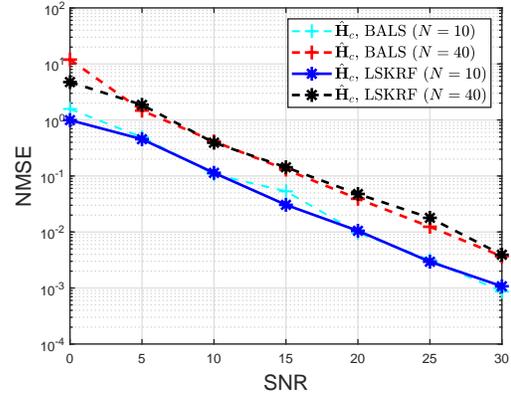}
\end{center}
	\vspace{-3ex}
\caption{NMSE of the estimated cascaded channel $\hat{\mathbf{H}}_{\textrm{c}}$.}
	\label{EFCOMP}
\end{figure}
\begin{figure}[!t]
\begin{center}
	\includegraphics[scale=0.5]{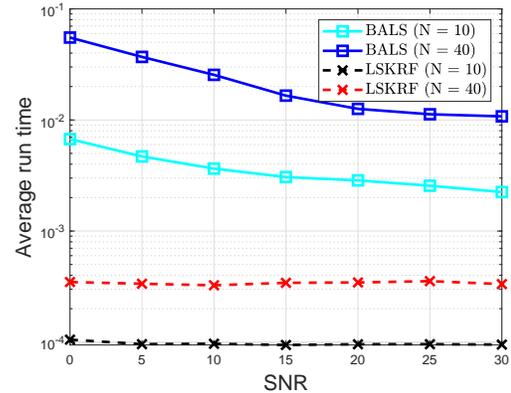}
\end{center}
	\vspace{-3ex}
\caption{Average runtime of LSKRRF and BALS algorithms.}
	\label{RTCOMP}
\end{figure}
All the results represent an average from $R=3000$ run Monte Carlo runs. Our simulations assume $T = 4$, $L = 2$, $K = 50$, and $M = 3$. To analyze the impact the number of IRS elements, we assume $N \in \{10, 40\}$. Figure \ref{GHCOMP} depicts the NMSE vs. SNR curves for the LSKRF and BALS algorithms. We can see that both algorithms provide satisfactory performances. The performances degrade as the number of IRS elements is increased, which is an expected result. 
In Figure \ref{EFCOMP}, the NMSE performance of the estimated cascaded channel $\hat{\mathbf{H}}_{\textrm{c}}\doteq \hat{\mathbf{G}}\hat{\mathbf{H}}^{\textrm{T}}$ is shown. The results follow the behavior of those of Figure \ref{GHCOMP}. 
In~Figure \ref{RTCOMP}, the average runtime of LSKRF and BALS methods are shown. We can note that BALS is more complex than LSKRF. The runtime of BALS grows faster then that of LSKRF with the increase of the number $N$ of IRS elements. On the other hand, as we pointed out earlier, BALS can operate under less restrictive choices for $K$ and $T$ in comparison with LSKRF. Hence, there is a tradeoff between complexity and operating conditions for the two channel estimation methods.

\section{Conclusion}
We have proposed two simple channel estimation methods for IRS-assisted MIMO systems based on PARAFAC modeling. The two proposed algorithms exhibit similar performances. The closed-form (LSKRF) method has a lower complexity but a more restrictive requirement on the training parameters $K$ and $T$, while the iterative (BALS) method, although being more computationally complex, can operate under more flexible choices for these parameters. 



\newpage

\bibliographystyle{IEEEtran}
\bibliography{IEEEexample}

\begin{thebibliography}{10}
\providecommand{\url}[1]{#1}
\csname url@samestyle\endcsname
\providecommand{\newblock}{\relax}
\providecommand{\bibinfo}[2]{#2}
\providecommand{\BIBentrySTDinterwordspacing}{\spaceskip=0pt\relax}
\providecommand{\BIBentryALTinterwordstretchfactor}{4}
\providecommand{\BIBentryALTinterwordspacing}{\spaceskip=\fontdimen2\font plus
\BIBentryALTinterwordstretchfactor\fontdimen3\font minus
  \fontdimen4\font\relax}
\providecommand{\BIBforeignlanguage}[2]{{%
\expandafter\ifx\csname l@#1\endcsname\relax
\typeout{** WARNING: IEEEtran.bst: No hyphenation pattern has been}%
\typeout{** loaded for the language `#1'. Using the pattern for}%
\typeout{** the default language instead.}%
\else
\language=\csname l@#1\endcsname
\fi
#2}}
\providecommand{\BIBdecl}{\relax}
\BIBdecl

\bibitem{Basar_2019}
E.~Basar, M.~D. Renzo, J.~D. Rosny, M.~Debbah, M.~S. Alouini, and R.~Zhang,
  ``Wireless communications through reconfigurable intelligent surfaces,''
  \emph{IEEE Access}, vol.~7, pp. 116\,753--116\,773, Aug 2019.

\bibitem{Survey_NOVO}
S.~Gong, X.~Lu, D.~T. Hoang, D.~Niyato, L.~Shu, D.~I. Kim, and Y.-C. Liang,
  ``Towards smart radio environment for wireless communications via intelligent
  reflecting surfaces: A comprehensive survey,'' 2019, [Online]. Available:
  https://arxiv.org/abs/1912.07794.

\bibitem{Liaskos2018ANW}
C.~Liaskos, S.~Nie, A.~Tsioliaridou, A.~Pitsillides, S.~Ioannidis, and I.~F.
  Akyildiz, ``A new wireless communication paradigm through software-controlled
  metasurfaces,'' \emph{IEEE Communications Magazine}, vol.~56, pp. 162--169,
  2018.

\bibitem{Jung2019}
M.~Jung, W.~Saad, Y.~Jang, G.~Kong, and S.~Choi, ``Performance analysis of
  large intelligent surfaces {(LISs)}: Asymptotic data rate and channel
  hardening effects,'' 2019, [Online]. Available:
  https://arxiv.org/abs/1810.05667.

\bibitem{Huang2019}
C.~Huang, A.~Zappone, G.~C. Alexandropoulos, M.~Debbah, and C.~Yuen,
  ``Reconfigurable intelligent surfaces for energy efficiency in wireless
  communication,'' \emph{IEEE Transactions on Wireless Communications},
  vol.~18, no.~8, p. 4157–4170, Aug 2019.

\bibitem{basar2019reconfigurable}
E.~Basar, ``Reconfigurable intelligent surface-based index modulation: A new
  beyond {MIMO} paradigm for {6G},'' 2019, [Online]. Available:
  https://arxiv.org/abs/1904.06704.

\bibitem{Tobias2019}
T.~L. Jensen and E.~D. Carvalho, ``On optimal channel estimation scheme for
  intelligent reflecting surfaces based on a minimum variance unbiased
  estimator,'' 2019, [Online]. Available: https://arxiv.org/abs/1909.09440v1.

\bibitem{ZHEN2019}
Z.~Q. He and X.~Yuan, ``Cascaded channel estimation for large intelligent
  metasurface assisted massive {MIMO},'' 2019,
  https://arxiv.org/abs/1905.07948v2.

\bibitem{NING2019}
B.~Ning, Z.~Chen, W.~Chen, Y.~Du, and J.~Fang, ``Channel estimation and hybrid
  beamforming for reconfigurable intelligent surfaces assisted {T}hz
  communications,'' December 2019, [Online]. Available:
  https://arxiv.org/pdf/1912.11662.

\bibitem{Yaoshen2019}
Y.~Cui and H.~Yin, ``An efficient {CSI} acquisition method for intelligent
  reflecting surface-assisted mmwave networks,'' December 2019, [Online].
  Available: https://arxiv.org/abs/1912.12076.

\bibitem{Emil:NOV2019}
E.~Bj{\"{o}}rnson, {\"{O}}.~{\"{O}}zdogan, and E.~G. Larsson, ``Intelligent
  reflecting surface vs. decode-and-forward: How large surfaces are needed to
  beat relaying?'' Nov 2019, [Online]. Available:
  http://arxiv.org/abs/1906.03949.

\bibitem{Harshman70}
R.~A. Harshman, ``Foundations of the {PARAFAC} procedure: Models and conditions
  for an explanatory multimodal factor analysis,'' \emph{UCLA Working Papers in
  Phonetics}, vol.~16, pp. 1--84, 1970.

\bibitem{Kolda2009}
T.~G. Kolda and B.~W. Bader, ``Tensor decompositions and applications,''
  \emph{SIAM Review}, vol.~51, no.~3, pp. 455--500, 2009.

\bibitem{CLdA09}
P.~Comon, X.~Luciani, and A.~L.~F. de~Almeida, ``{Tensor decompositions,
  alternating least squares and other tales},'' \emph{{Journal of
  Chemometrics}}, vol.~23, no. 7-8, pp. 393--405, 2009.

\bibitem{Almeida2016}
A.~L.~F. de~Almeida, G.~Favier, J.~P. C.~L. da~Costa, and J.~C.~M. Mota,
  ``{Overview of tensor decompositions with applications to communications},''
  in \emph{{Signals and Images: Advances and Results in Speech, Estimation,
  Compression, Recognition, Filtering, and Processing}}, R.~Coelho,
  V.~Nascimento, R.~de~Queiroz, J.~Romano, and C.~Cavalcante, Eds.\hskip 1em
  plus 0.5em minus 0.4em\relax {CRC-Press}, Jan. 2016, no. Chapter 12, pp.
  325--356.

\bibitem{Sidiropoulos2017}
N.~D. {Sidiropoulos}, L.~{De Lathauwer}, X.~{Fu}, K.~{Huang}, E.~E.
  {Papalexakis}, and C.~{Faloutsos}, ``Tensor decomposition for signal
  processing and machine learning,'' \emph{IEEE Transactions on Signal
  Processing}, vol.~65, no.~13, pp. 3551--3582, July 2017.

\bibitem{jensen2019optimal}
T.~L. Jensen and E.~D. Carvalho, ``An optimal channel estimation scheme for
  intelligent reflecting surfaces based on a minimum variance unbiased
  estimator,'' 2019, [Online]. Available: https://arxiv.org/pdf/1909.09440.

\bibitem{Kibangou2009}
A.~Y. {Kibangou} and G.~{Favier}, ``Non-iterative solution for parafac with a
  toeplitz matrix factor,'' in \emph{2009 17th European Signal Processing
  Conference}, Aug 2009, pp. 691--695.

\bibitem{Roemer2010}
F.~{Roemer} and M.~{Haardt}, ``Tensor-based channel estimation and iterative
  refinements for two-way relaying with multiple antennas and spatial reuse,''
  \emph{IEEE Transactions on Signal Processing}, vol.~58, no.~11, pp.
  5720--5735, Nov 2010.

\bibitem{golub13}
G.~H. Golub and C.~F. van Loan, \emph{Matrix Computations}, 4th~ed.\hskip 1em
  plus 0.5em minus 0.4em\relax John Hopkins University Press, 2013.

\end{thebibliography}
\end{document}